\documentclass[12pt]{iopart}
\usepackage{graphicx}
\usepackage{dcolumn}
\usepackage{bm}
\usepackage{latexsym}
\begin{document}
\paper{Competition of coarsening and shredding of clusters in a
driven diffusive lattice gas}
\author{Ambarish Kunwar$^1$\footnote{E-mail: ambarish@iitk.ac.in},
 Debashish Chowdhury$^2$\footnote{E-mail: debch@iitk.ac.in}, 
 Andreas Schadschneider$^3$\footnote{E-mail: as@thp.uni-koeln.de} and 
Katsuhiro Nishinari$^4$\footnote{E-mail: tknishi@mail.ecc.u-tokyo.ac.jp}}

\address{$^1$Department of Physics, Indian Institute of Technology, 
Kanpur 208016, India.}

\address{$^2$Department of Physics, Indian Institute of Technology, 
Kanpur 208016, India.}

\address{$^3$Institute for Theoretical Physics, University of Cologne, 
D-50937 K\"oln, Germany.}

\address{$^4$Department of Aeronautics and Astronautics, Faculty of 
Engineering, University of Tokyo, Hongo, Bunkyo-ku, Tokyo 113-8656, Japan.}

\begin{abstract} 
We investigate a driven diffusive lattice gas model with two oppositely 
moving species of particles. The model is motivated by bi-directional 
traffic of ants on a pre-existing trail. A third species, corresponding 
to pheromones used by the ants for communication, is not conserved and 
mediates interactions between the particles. Here we study the 
spatio-temporal organization of the particles. In the uni-directional 
variant of this model it is known to be determined by the formation and 
coarsening of ``loose clusters''. For our bi-directional model, we show 
that the interaction of oppositely moving clusters is essential. 
In the late stages of evolution the cluster size oscillates because of a 
competition between their `shredding' during encounters with oppositely 
moving counterparts and subsequent "coarsening" during collision-free 
evolution. We also establish a nontrivial dependence of the spatio-temporal 
organization on the system size.
\end{abstract}
\section{Introduction}
Systems of interacting driven particles are of current interest in 
statistical physics for understanding the fundamental principles of 
spatio-temporal organization far from equilibrium as well as for  
possible wide ranging practical applications \cite{sz,schutz}. 
The simplest model of this type with only one species of particles 
is the so-called totally asymmetric simple exclusion process (TASEP).
In the TASEP, a particle is picked up randomly and moved forward by 
one lattice spacing, with the hopping probability $q_{+}$, provided 
the target site on the one-dimensional lattice is empty.

\begin{figure}[ht] 
\begin{center} 
\includegraphics[width=0.58\columnwidth]{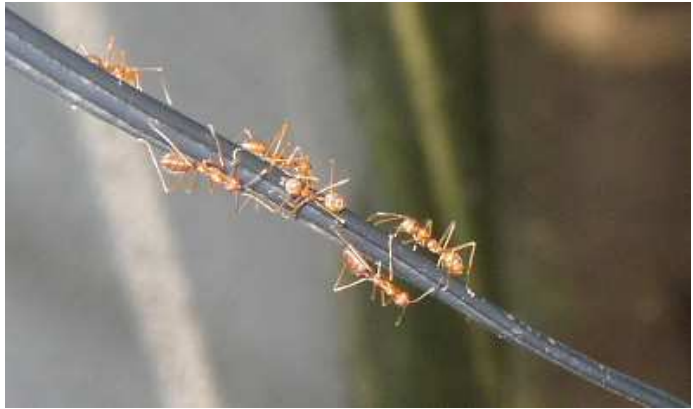} 
\end{center} 
\caption{A snapshot of an ant-trail on a hanging cable. It can be 
regarded as strictly one-dimensional. But, nevertheless, traffic 
flow in opposite directions is possible as two ants, which face each 
other on the upper side of the cable, can exchange their positions 
if one of them, at least temporarily, switches over to the lower 
side of the cable.} 
\label{fig-antphoto}
\end{figure} 
\noindent
Most of the earlier models with more than one species of particles 
\cite{mukamel,evans1,evans2,arndt,rajewsky,lahiri,naim,popkov}  
addressed mainly the questions on the (im)possibility of spontaneous 
symmetry breaking, phase separation, etc.\ in {\it one-dimensional} 
stochastic driven systems. Here we extend the two-species 
models like, for example, ABC model \cite{evans1,evans2} and AHR model 
\cite{arndt} to develop a {\it three-species} model \cite{foot}. The 
density of the new species introduced in our model, 
which corresponds to the pheromones secreted by the
ants for communication, 
is not conserved by the dynamics of the system and its coupling to the
two other conserved variables leads to uncommon features of the
spatio-temporal organisations of the conserved particles.
For a special choice of the model parameters our model becomes 
identical to the AHR model classes (see Sec.~3).

Our model is motivated by a real-life situation where {\em bi-directional} 
ant-traffic in an effectively one-dimensional system is observed.
Fig.~\ref{fig-antphoto} shows a hanging cable which is part of an ant-trail.
Ants prefer moving on the upper side of the cable, which would make 
the motion strictly one-dimensional. If ants moving in
opposite directions meet head-on, after short ``negotiations'' one
of them would switch to the lower part of the cable temporarily
in order to make passing possible. Due to the ``negotiations''
this process leads to slowing down of both the ants. We capture 
this entire process by an exchange of the positions of the two 
ants with a sufficiently low probability which corresponds to a 
slower movement of each of them as compared to a freely moving ant. 
The mathematical formulation of the model in terms of such exchange 
probabilities is very similar to the formulations of the class of 
one-dimensional stochastic models of driven-diffusive lattice gases 
to which the AHR and ABC models belong.

The number of particles leaving a site per unit 
time is called the flux or current; the flux-versus-density relation 
is often referred to as the {\it fundamental diagram}. We study 
interplay of spatio-temporal organization and flow (fundamental diagram) 
in the system by a combination of analytical arguments and extensive 
computer simulations. 


\section{The model}
\begin{figure}
\begin{center}
\begin{tabular}{| c | c | c |}
\hline
initial & final & rate\\
\hline
RL & RL & $1-K$\\
   & LR & $K$\\
\hline
RP & RP & $(1-f)(1-Q)$\\
   & R0 & $f(1-Q)$\\
   & 0R & $fQ$\\
   & PR & $(1-f)Q$\\
\hline
R0 & R0 & $1-q$\\
   & 0R & $fq$\\
   & PR & $(1-f)q$\\
\hline
PR & PR & $1-f$\\
   & 0R & $f$\\
\hline
P0 & P0 & $1-f$\\
   & 00 & $f$\\
\hline
PP & PP & $(1-f)^2$\\
   & P0 & $f(1-f)$\\
   & 0P & $f(1-f)$\\
   & 00 & $f^2$\\
\hline
\end{tabular}
\caption{Nontrivial transitions and their transition rates. Transitions
from initial states $PL$, $0L$ and $0P$ are not listed. They can be
obtained from those for $LP$, $L0$ and $P0$, respectively, by replacing
$R\leftrightarrow L$ and, then, taking the mirror image.}
\label{fig-updating}
\end{center}
\end{figure}
\noindent
\noindent
In our model the right-moving (left-moving) particles, represented by 
$R$ ($L$), are never allowed to move towards left (right); these two 
groups of particles are the analogs of the outbound and nest-bound 
ants in a {\it bi-directional} traffic on the same trail. Thus, no 
U-turn is allowed. In addition to the TASEP-like hopping of the 
particles onto the neighboring vacant sites in the respective directions 
of motion, the $R$ and $L$ particles on nearest-neighbour sites and 
facing each other are allowed to exchange their positions, i.e., the 
transition $ RL~{\to}~LR$ takes place, with the 
probability $K$. This might be considered as a minimal model for the 
motion of ants on a hanging cable as shown in Fig.~\ref{fig-antphoto}. 
When a outbound ant and a nest-bound ant face each other on the upper 
side of the cable, they slow down and, eventually, pass each other 
after one of them, at least temporarily, switches over to the 
lower side of the cable. Similar observations have been made 
for normal ant-trails where ants pass each other after turning by a 
small angle to avoid head-on collision \cite{couzin,burd2}. In our 
model, as commonly observed in most real ant-trails, none of the ants 
is allowed to overtake another moving in the same direction. 

\noindent
Ants drop a chemical (generically called {\it pheromone}) on the 
substrate as they move forward \cite{wilson}. They can follow the 
trail by picking up the ``smell'' of the trail pheromone provided 
the pheromone evaporates sufficiently slowly with time. We now 
introduce a third species of particles, labelled by the letter $P$, 
which are intended to capture the essential features of pheromone. 
The $P$ particles are deposited on the lattice by the $R$ and $L$ 
particles when the latter hop out of a site; an existing $P$ 
particle at a site disappears when a $R$ or $L$ particle arrives 
at the same location. The $P$ particles cannot hop but can {\it 
evaporate}, with a probability $f$ per unit time, independently 
from the lattice. None of the lattice sites can accommodate more 
than one particle at a time. From now onwards, we shall refer to 
this model as the $PRL$ model.
\noindent
The state of the system is updated in a {\it random-sequential} manner. 
Because of the {\it periodic boundary conditions}, the densities of the $R$ 
and the $L$ particles are conserved. In contrast, the density of the $P$ 
particles is a non-conserved variable. The distinct initial states and 
the corresponding final states for pairs of nearest-neighbor sites are 
shown in fig.\ref{fig-updating} together with the respective transition 
probabilities.

\begin{figure}[tb]
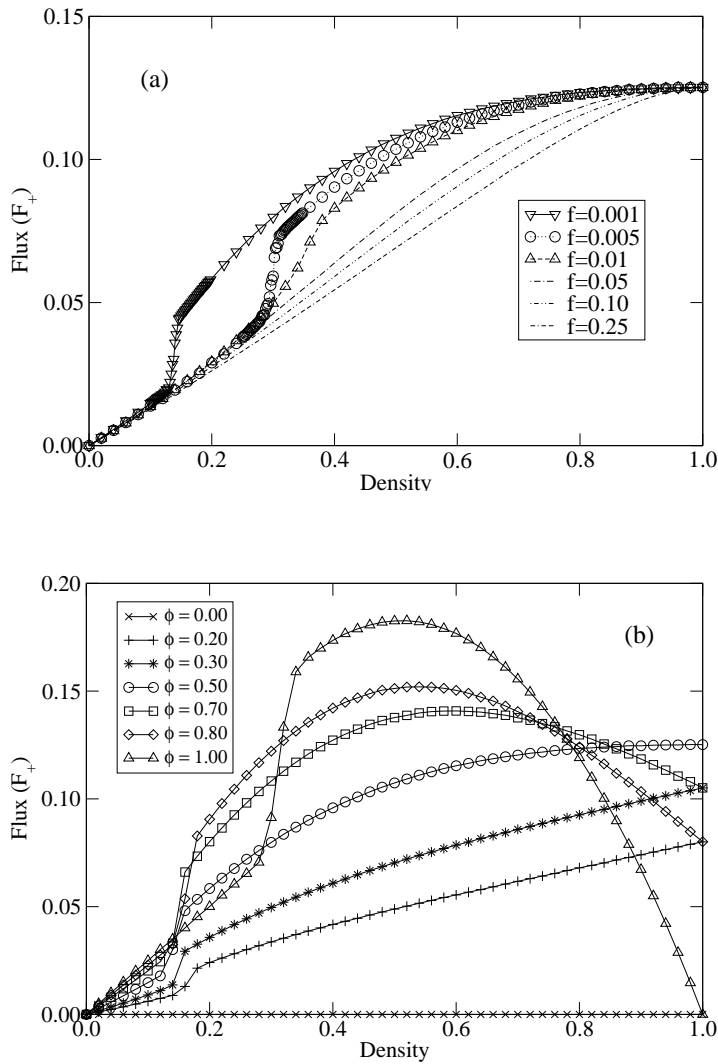

\vspace{0.75cm}
\begin{center}
\includegraphics[width=0.6\textwidth]{fig2a.eps} 
\hspace{1cm}\\
\vspace{1cm}
\includegraphics[width=0.6\textwidth]{fig2b.eps} 
\end{center}
\caption{The fundamental diagrams in the steady-state of the PRL 
model for several different values of (a) $f$ (for $\phi = 0.5$) 
and (b) $\phi$ (for $f = 0.001$). The other common 
parameters are $Q = 0.75, q = 0.25$, $K = 0.5$ and $M = 1000$. 
}
\label{fig-prlfd}
\end{figure}

\begin{figure}[tbp]
\begin{center}
\includegraphics[width=0.55\columnwidth]{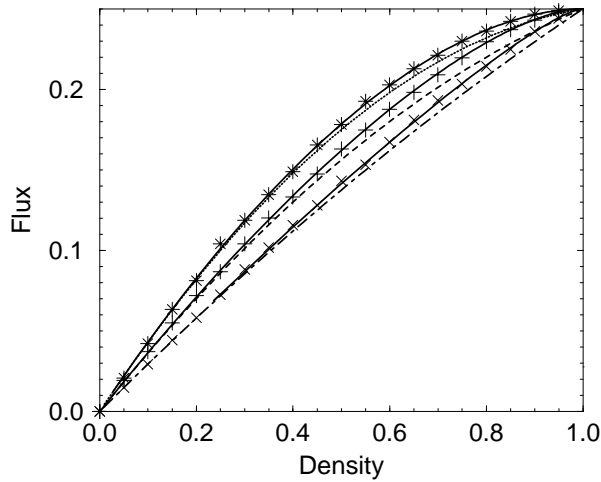}
\end{center}
\caption{Fundamental diagram in the special case $Q = q = q_H$.
The continuous curves, marked by $\ast$, $+$ and $\times$, are the
exact results corresponding to $q_H = 0.90, 0.75, 0.60$, respectively.
The corresponding HMFA results have been shown by the dotted, dashed
and dashed-dotted curves, respectively. The points marked by
$\ast$, $+$ and $\times$ have been obtained from computer simulations.
}
\label{fig-hmfa}
\end{figure}

\begin{figure}[tb]
\vspace{0.55cm}
\begin{center}
\includegraphics[width=0.55\textwidth]{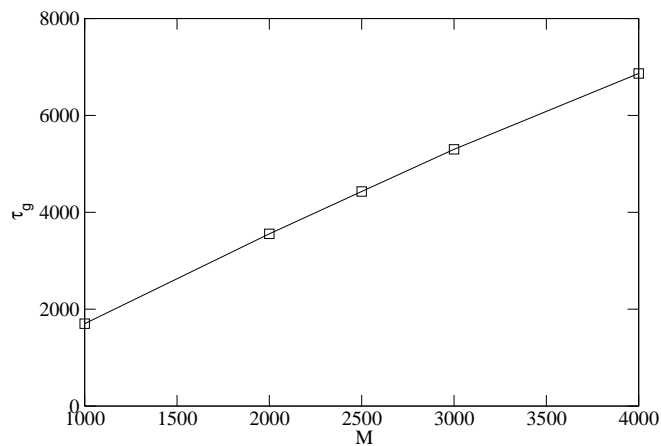} 
\end{center}
\caption{Time gap $\tau_g$ between successive collisions as a function of 
system size M for $Q = 0.75$, $q = 0.25$, $K=0.5$, $f = 0.005$, $c = 0.2$ and
$\phi = 0.3$}
\label{fig-tauvsl}
\end{figure}

\begin{figure}[tb]
\begin{center}
\includegraphics[width=0.475\textwidth]{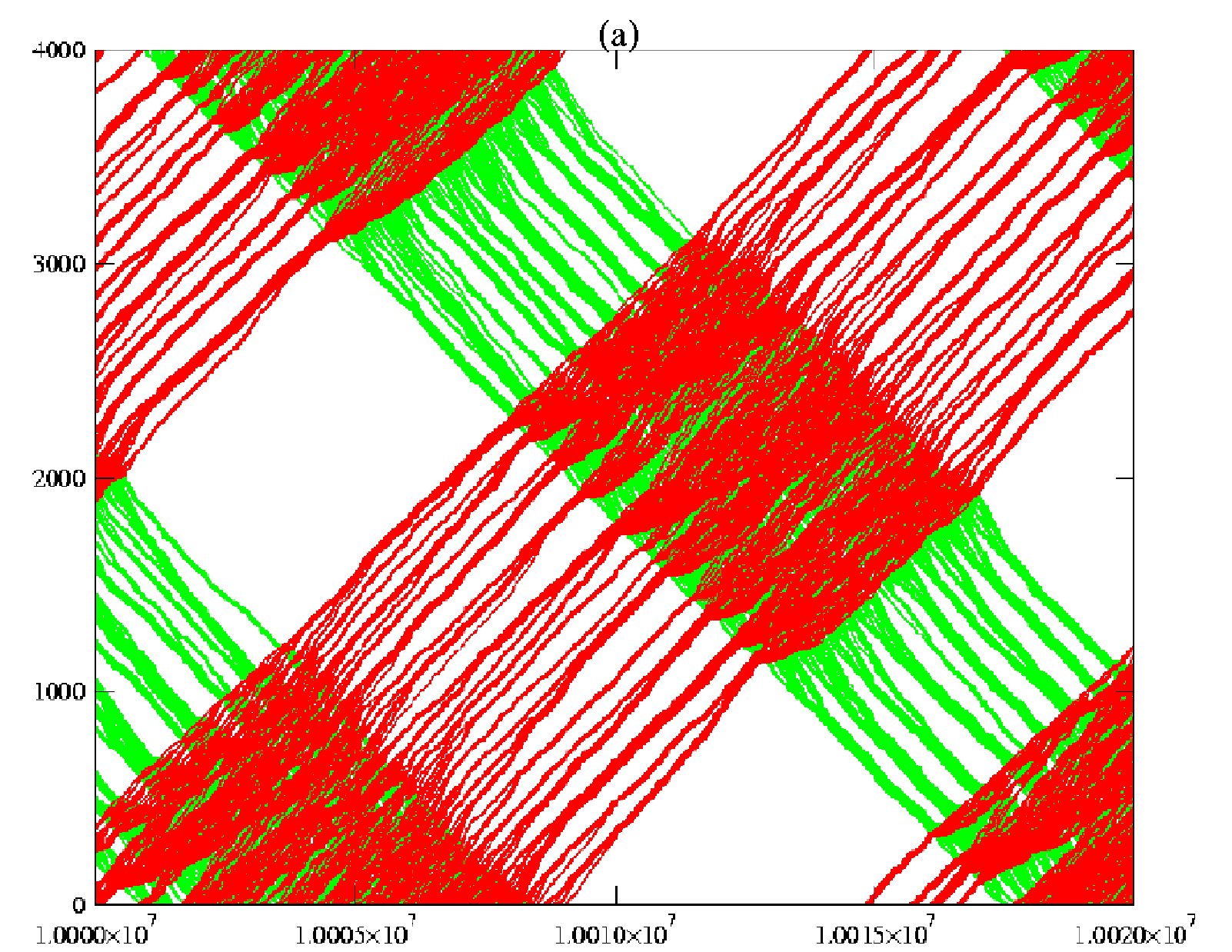} 
\includegraphics[width=0.475\textwidth]{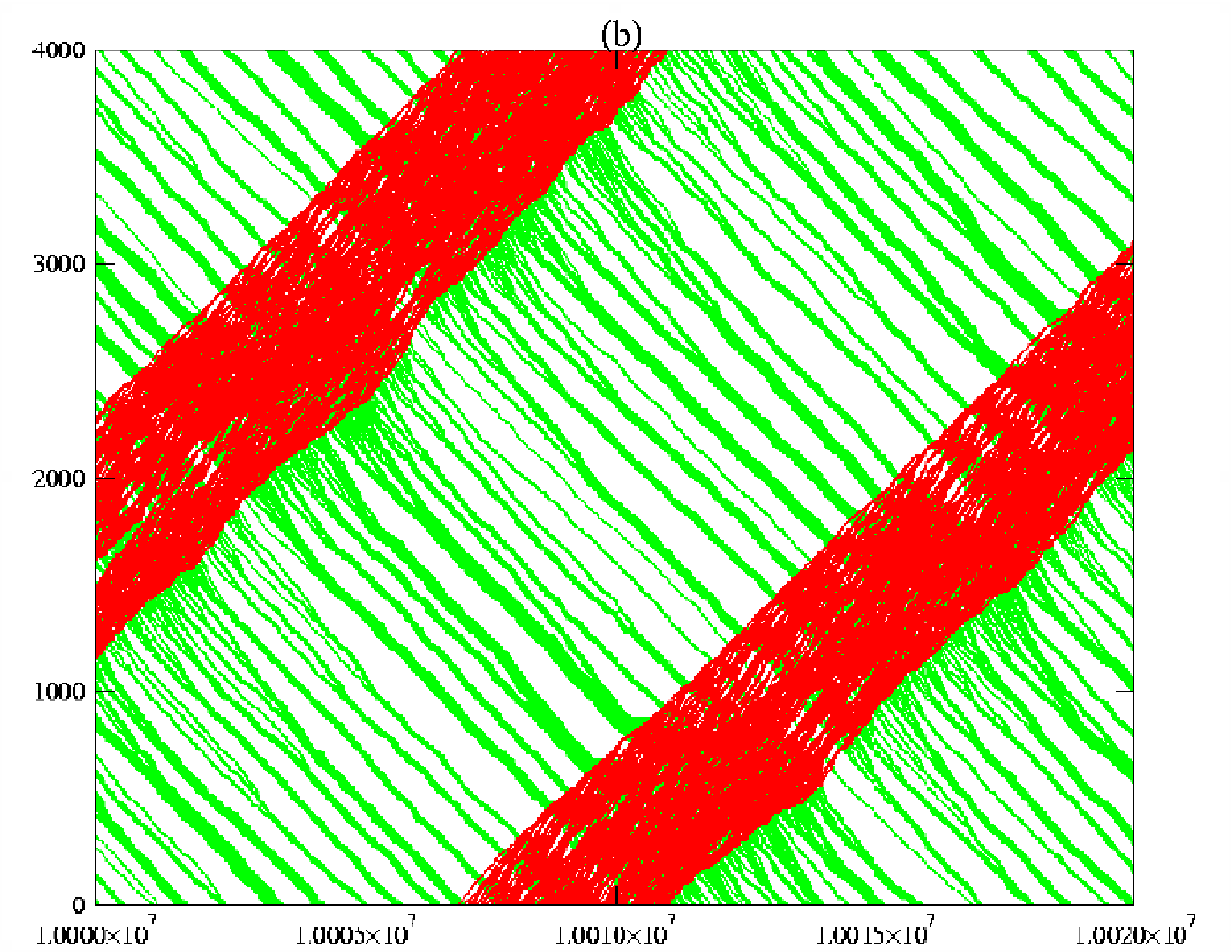}
\includegraphics[width=0.475\textwidth]{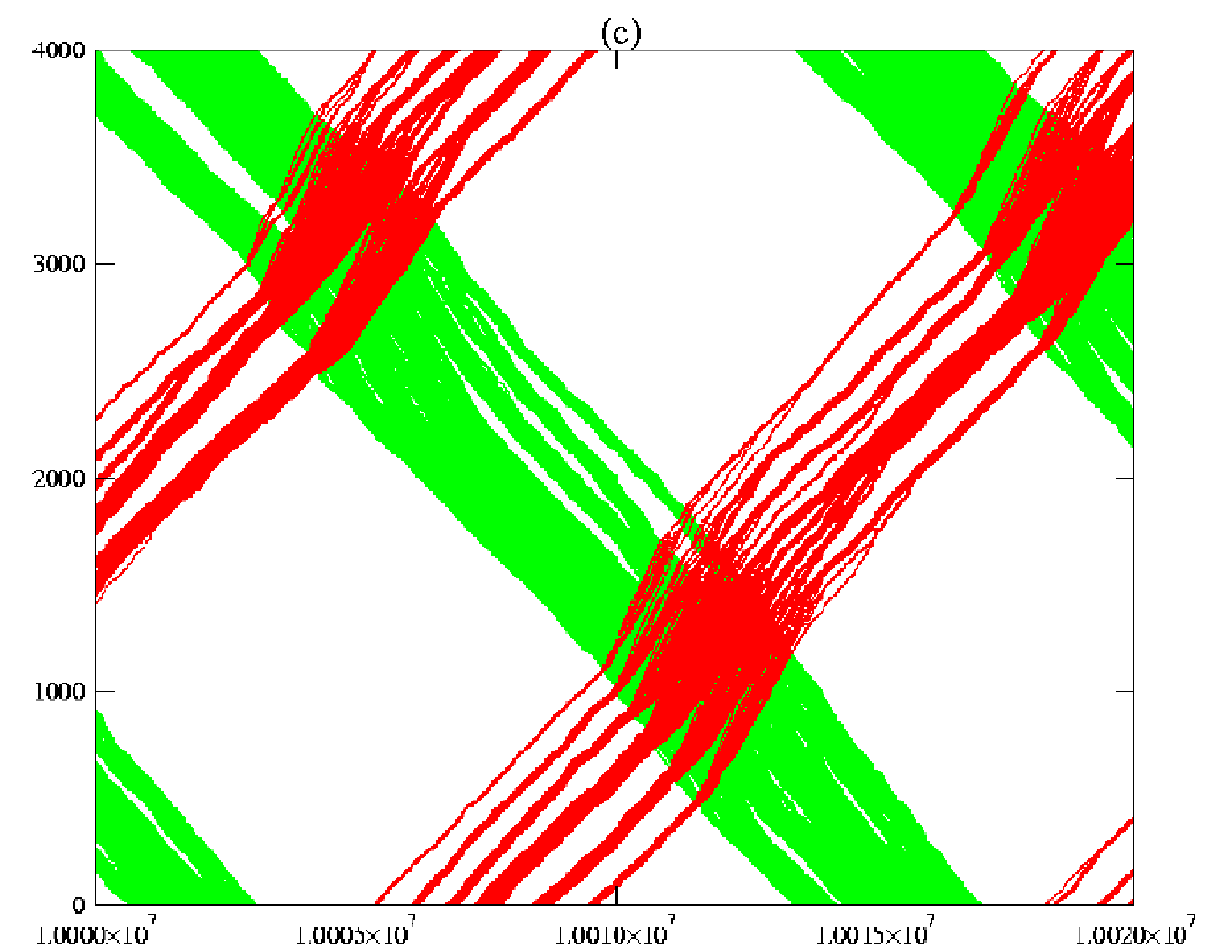}
\end{center}
\caption{Space-time plot of the PRL model for  
$Q = 0.75$, $q = 0.25$, $f = 0.005$, $M = 4000$, $c = 0.2$ and 
(a) $\phi = 0.5$, $K = 0.2$, (b) $\phi = 0.3$, $K = 0.2$, 
(c) $\phi = 0.3$, $K = 0.5$. The red and green dots represent 
the right-moving and left-moving ants, respectively. }
\label{fig-prlst}
\end{figure}

\begin{figure}[tb]
\begin{center}
\vspace{0.5cm}
\includegraphics[angle=-90,width=0.485\textwidth]{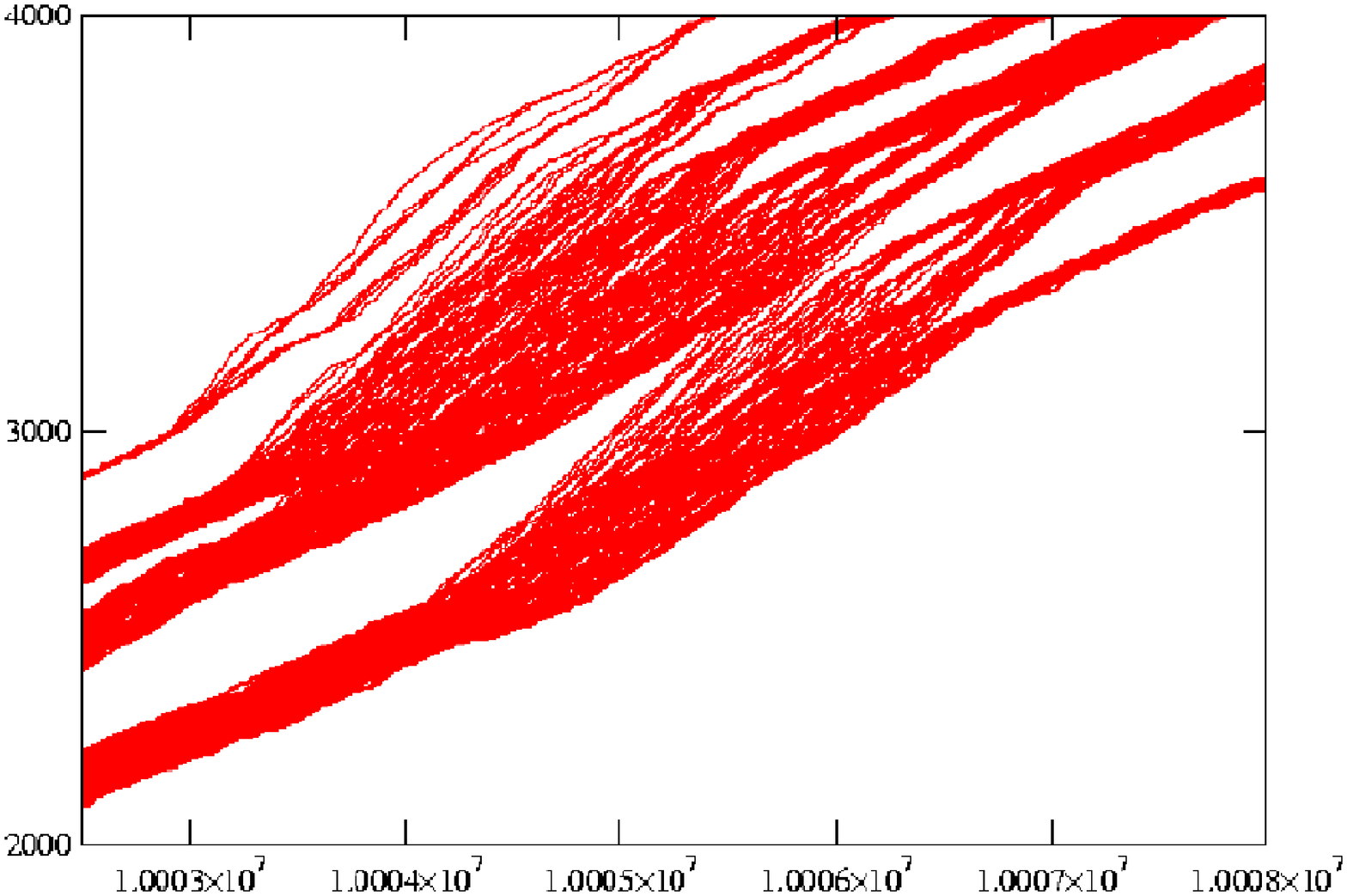}
\includegraphics[angle=-90,width=0.485\textwidth]{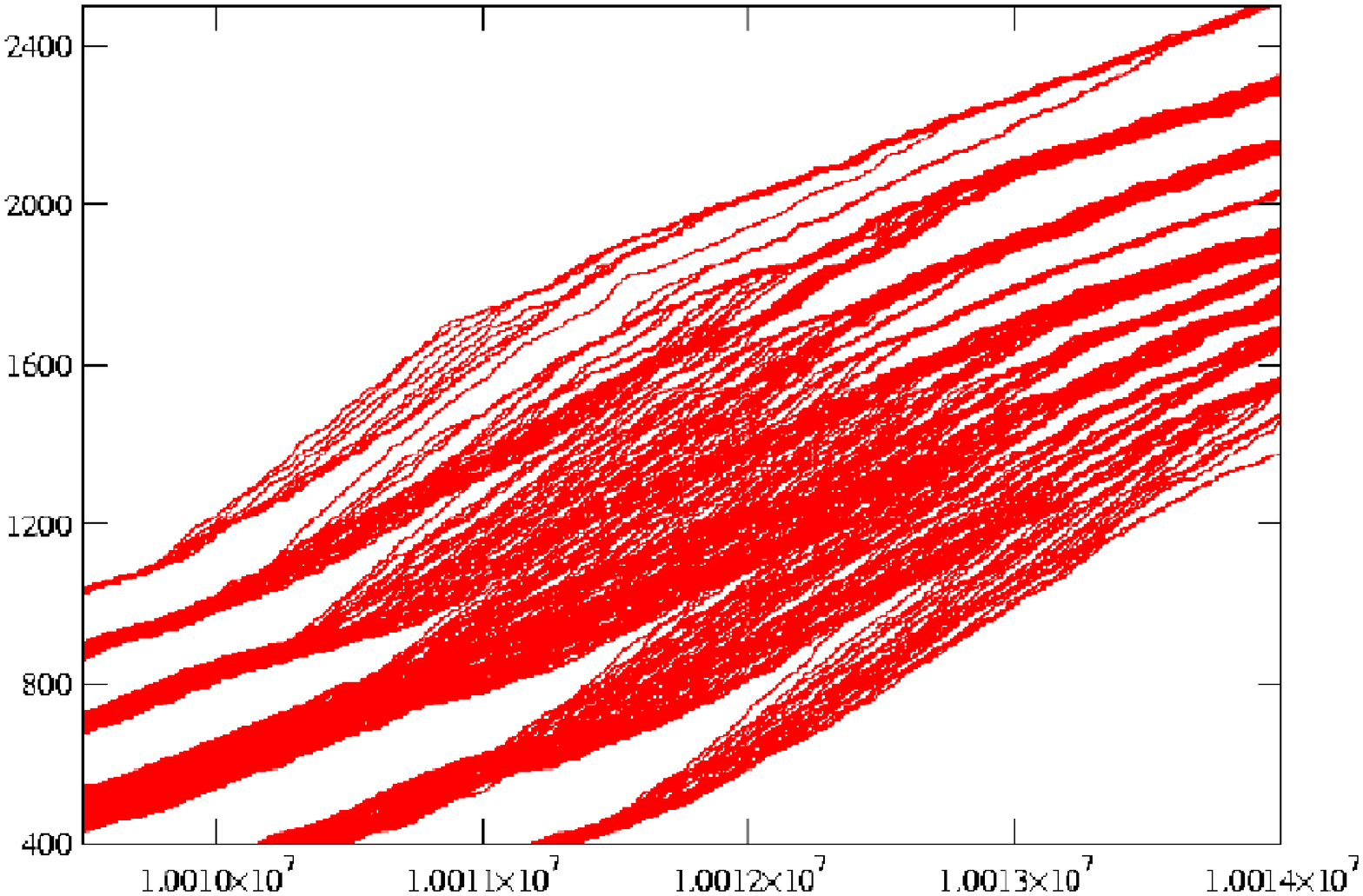}
(a) \hspace{7cm}(b)
\end{center}
\caption{Magnified view of the first and second collision area respectively of Fig. \ref{fig-prlst}(c)}
\label{fig-collision}
\end{figure}
\begin{figure}[tb]
\begin{center}
\vspace{0.5cm}
\includegraphics[width=0.475\textwidth]{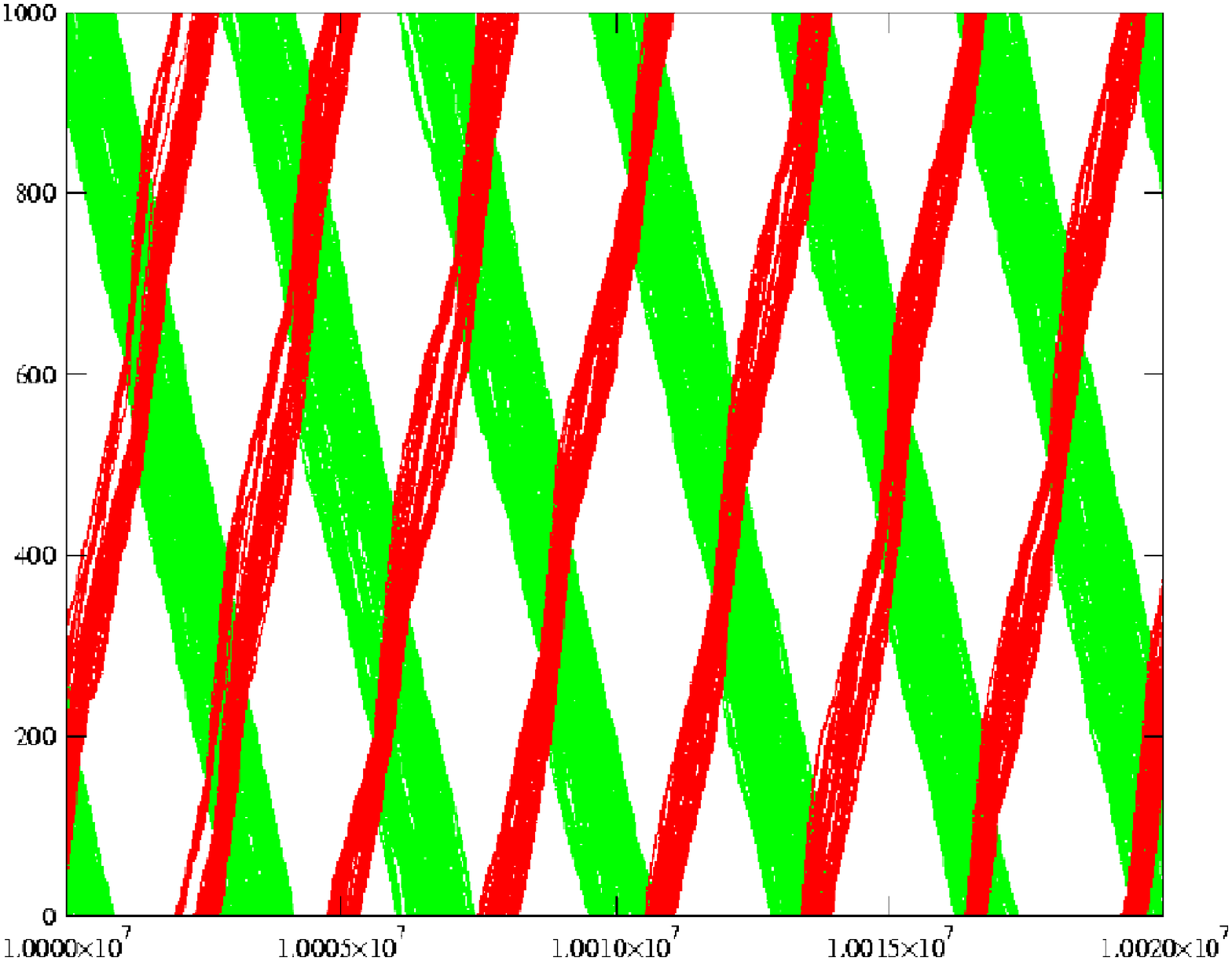} 
\includegraphics[width=0.475\textwidth]{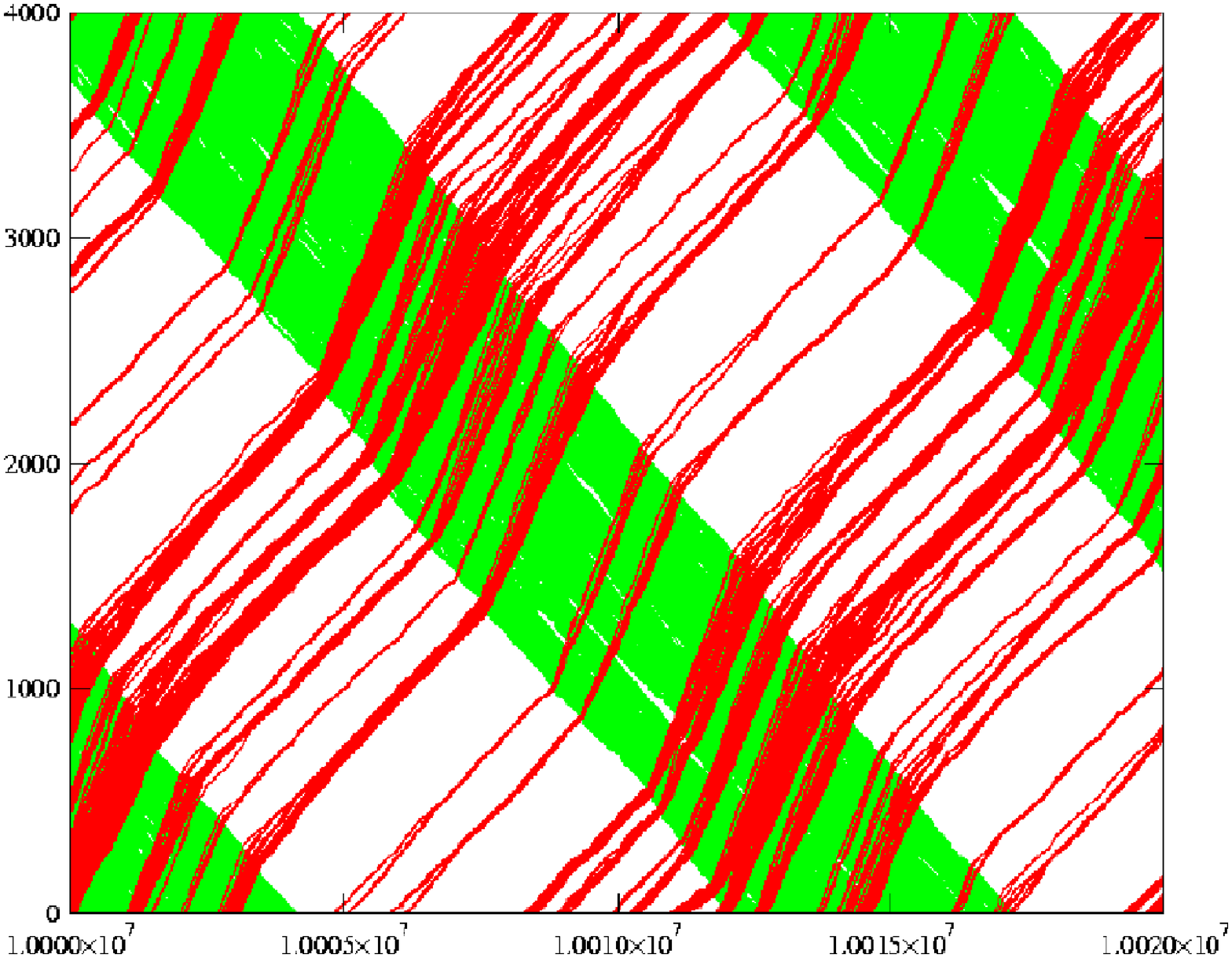}
(a) \hspace{7cm} (b) 
\end{center}
\caption{Space-time plot of the PRL model for  $Q = 0.50$, $q = 0.25$, 
$f = 0.005$, $c = 0.2$, $\phi = 0.3$, $K = 1.0$ and (a) $M = 1000$, 
(b) $M = 4000$. The red and green dots represent the right-moving and 
left-moving ants, respectively. 
}
\label{fig-prlsize}
\end{figure}

\section{Results}
\noindent
Suppose $N_+$ and $N_- = N - N_+$ are the total numbers of $R$ and $L$ 
particles, respectively. For a system of length $M$ the corresponding 
densities are $c_{\pm} = N_{\pm}/M$  with the total density 
$c = c_+ + c_- = N/M$. Of the $N$ particles, a fraction 
$\phi = N_{+}/N = c_{+}/c$ are of the type $R$ while the remaining 
fraction $1-\phi$ are $L$ particles. The corresponding fluxes are 
denoted by $F_{\pm}$. In both the limits $\phi = 1$ and $\phi = 0$ this 
model reduces to our earlier model \cite{cgns,ncs}, motivated by 
uni-directional ant-traffic, which is closely related to the bus-route 
models \cite{loan,cd} and the models of pedestrian dynamics 
\cite{helbing2,schad}. 
\noindent
One unusual feature of this PRL model is that the flux does {\it not} 
vanish in the {\it dense-packing} limit $c \rightarrow 1$. In fact, 
in the {\it full-filling} limit $c = 1$, the {\it exact} non-vanishing 
flux $F_+ = K c_+ c_- = F_-$ at $c_++c_- = c = 1$ arises only from the 
exchange of the $R$ and $L$ particles, {\it irrespective of the 
magnitudes of} $f, Q$ and $q$.

\noindent
In the special case $Q = q = q_H$ the hopping of the ants become 
independent of pheromone. This special case of the PRL model is 
identical to the AHR model \cite{arndt} with $q_- = 0 = \kappa$. 
A simple homogeneous mean-field approximation (HMFA) yields the estimates 
\begin{eqnarray}
F_{\pm} \simeq c_{\pm} \biggl[ q_H (1-c) + K c_{\mp} \biggr] 
\label{eq-mf}
\end{eqnarray} 
{\it irrespective of} $f$, for the fluxes $F_{\pm}$ at any arbitrary $c$. 
On the other hand, the exact
expression for the flux of the AHR model, parametrized \cite{rajewsky}
by a rapidity variable $\xi \in [0,\frac{1}{(1-a)^2}]$, is given by:
\begin{eqnarray}
J(\xi) &=& \frac{2a^2\xi}{\cal N},\\
\rho(\xi) &=& 2\frac{a(1+a)\xi\left[(1+a)\sqrt{1+4a\xi}-(1-a)\right]}{
            {\cal N}\sqrt{1+4a\xi}}
\label{eq-4}
\end{eqnarray}
where
\begin{equation}
{\cal N} = 1+a^2+2a(1+a)^2\xi - (1-a^2) \sqrt{1+4a\xi}.
\end{equation}
for $1/2 \leq q_H \leq 1$, where $a=\frac{1-q_H}{q_H}$ and the unit of
elementary time scale has been set by choosing $K = 1$. A comparison of
the equation (\ref{eq-mf}) (with $K = 1$) and the exact result (\ref{eq-4})
in Fig.~\ref{fig-hmfa} shows that the flux in the HMFA, although an
underestimate, is a reasonably good approximation for all $q_H \geq 1/2$.
Deviation from the exact values for $q_H < 1/2$ indicates the presence of 
stronger correlations at smaller values of $q_H$.
\noindent
For the generic case $q \neq Q$, the flux in the PRL model depends on 
the evaporation rate $f$ of the $P$ particles. In Fig.~\ref{fig-prlfd} 
we plot the fundamental diagrams for wide ranges of values of $f$ 
(in Fig.~\ref{fig-prlfd}(a)) and $\phi$ (in Fig.~\ref{fig-prlfd}(b)), 
corresponding to one set of hopping probabilities. 
First, note that the data in Figs.~\ref{fig-prlfd} are consistent 
with the physically expected value of 
$F_{\pm}(c = 1) = K c_{+} c_{-}$, because in the dense packing 
limit only the exchange of the oppositely moving particles 
contributes to the flux. Moreover, the sharp rise of the flux 
over a narrow range of $c$ observed in both Fig.~\ref{fig-prlfd} (a) 
and (b) arise from the nonmonotonic variation of the average speed 
with density, an effect which was also observed in our earlier model 
for uni-directional ant traffic \cite{cgns,ncs}. 
This nonmonotonicity was shown to be a consequence of the
formation of so-called {\em loose clusters}. These are regions
in space where the particle density $c_{\rm lc}$ is larger than the 
average density $c$, but not maximal, i.e.\ $c < c_{\rm lc} < 1$.
These loose cluster dominate the behaviour at intermediate particle
densities $c$ and small evaporation rates $f$ where they are formed due 
to effectively longer-ranged attractive interactions introduced by
the pheromones.

\noindent
In the special limits $\phi = 0$ and $\phi = 1$, over a certain regime 
of density (especially at small $f$), the particles are known 
\cite{cgns,ncs} to form ``loose'' (i.e., non-compact) clusters 
\cite{ncs} which are defined to be regions of the system with a density 
$c_{\ell c}$ that is larger than the average global density, i.e., 
$c < c_{\ell c} < 1$. If the system evolves from a random initial 
condition at $t = 0$, then during coarsening of the cluster, its size 
$R(t)$ at time $t$ is given by $R(t) \sim t^{1/2}$ \cite{loan,cd}. 
Therefore, in the absence of encounter with oppositely moving particles, 
$\tau_{\pm}$, the coarsening time for the right-moving and left-moving 
particles would grow with system size as $\tau_{+} \sim \phi^2 M^2$ and 
$\tau_{-} \sim (1-\phi)^2 M^2$. 

\noindent
In the PRL model {\it with periodic boundary conditions}, the 
oppositely moving `loose'' clusters ``collide'' against each 
other periodically. Let us define $\tau_g$ to be the time gap between 
the successive collisions of the two clusters which is the time interval 
between the end of one collision and the beginning of the next collision. 
It is straightforward to see that $\tau_g$ increases {\it linearly} with 
the system size following $\tau_g \sim M$; we have verified this scaling 
relation numerically as in shown in Fig.~\ref{fig-tauvsl}. 
During a collision each loose cluster ``{\it shreds}'' the 
oppositely moving cluster; both clusters shred the other equally 
if $\phi = 1/2$ (Fig.~\ref{fig-prlst}(a)). However, for all 
$\phi \neq 1/2$, the minority cluster suffers more severe 
shredding than that suffered by the majority cluster 
(Fig.~\ref{fig-prlst}(b)) because each member of a cluster 
contributes in the shredding of the oppositely moving cluster.

\noindent
Fig.~\ref{fig-collision}(a) and \ref{fig-collision}(b) show magnified view
of the first and second collision area respectively in the space-time
plot Fig.~\ref{fig-prlst}(c) where only right moving ants are present.
Fig.~\ref{fig-collision}(a) and \ref{fig-collision}(b) demonstrate
that the phenomenon of shredding is such that the total number of clusters, 
their sizes as well as their individual velocities just after the collision 
are usually different from those of the clusters before the collision. 
But, at present,  we do not have any theory to predict the changes in 
these properties caused by the collision.
\begin{figure}[tbp]
\vspace{0.7cm}
\begin{center}
\includegraphics[width=0.5\textwidth,angle=0]{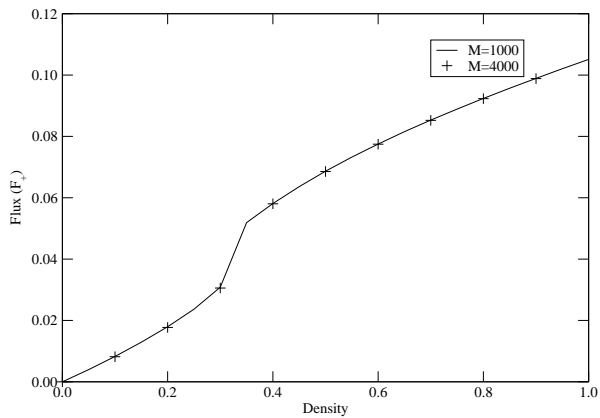}
\end{center}
\caption{Fundamental diagrams for the system size $M = 1000$ and $M = 4000$;
common parameters being $c = 0.2$, $\phi = 0.3$,  $Q = 0.75$, $q = 0.25$, 
$K = 0.50$ and $f = 0.005$}
\label{fundamental}
\end{figure}

\noindent
In small systems the ``shredded'' clusters get opportunity 
for significant re-coarsening before getting shredded again in 
the next encounter with the oppositely moving particles. But, in 
sufficiently large systems, shredded appearance of the clusters 
persists. This is demonstrated clearly by the space-time plots 
for two different system sizes in Fig.~\ref{fig-prlsize}. However, 
we observed practically no difference in the fundamental diagrams 
for $M = 1000$ and $M = 4000$ (see Fig.~\ref{fundamental}).
\begin{figure}[tb]
\vspace{0.7cm}
\begin{center}
\includegraphics[width=0.5\textwidth]{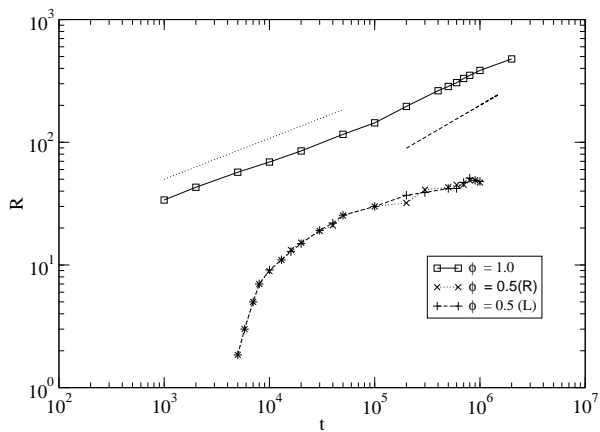} 
\end{center}
\caption{Average size of the cluster $R$ plotted against time $t$ 
for $\phi = 1.0$, and $\phi = 0.5$, both for the same total 
density $c = 0.2$; the other common parameters being 
$Q = 0.75$, $q = 0.25$, $K = 0.50$, $f = 0.005$, $M = 4000$.
Dotted line corresponds to a slope of $t^{1/3}$ and dashed line
corresponds to a slope of $t^{1/2}$. }  
\label{fig-rvst}
\end{figure}
Following the methods of ref.~\cite{cd}, we have computed $R(t)$ 
starting from random initial conditions. The data corresponding to 
$\phi = 1$ are consistent with the asymptotic growth law 
$R(t) \sim t^{1/2}$. In the begining $R(t)$ grows as $t^{1/3}$ 
(dotted line in Fig. \ref{fig-rvst} corresponds to a slope of 
$t^{1/3}$) however in the later stage it grows as $t^{1/2}$ (dashed 
line in Fig.~\ref{fig-rvst} corresponds to a slope of $t^{1/2}$). 
In sharp contrast, for $\phi = 0.5$, $R(t)$ saturates to a much 
smaller value that is consistent with highly shredded appearance of 
the clusters in Fig.~\ref{fig-prlst}(a).  

\noindent
Thus, coarsening and shredding phenomena compete against each 
other and this competition determines the overall spatio-temporal 
pattern. Therefore, in the late stage of evolution, the system 
settles to a state where, because of alternate occurrence of 
shredding and coarsening, the typical size of the clusters varies 
periodically. Moreover, comparing Fig.~\ref{fig-prlst}(b) and 
Fig.~\ref{fig-prlst}(c), we find that, for given $c$ and $\phi$, 
increasing $K$ leads to  sharper {\it speeding up} of the clusters 
during collision so long as $K$ is not much smaller than $q$. Both 
the phenomena of shredding and speeding during collisions of the 
oppositely moving loose clusters arise from the fact that, during 
such collisions, the dominant process is the exchange of positions, 
with probability $K$, of oppositely-moving ants that face each other. 

\section{Conclusions}
\noindent
The $PRL$ model reported in this paper, is motivated by bi-directional 
ant traffic. In a spatially constrained situation, e.g., on a hanging
cable (Fig.\ref{fig-antphoto}), such a single-lane model is adequate, 
whereas otherwise a full two-lane model \cite{johnetal} is required. 

The main effect of the new species of particles $P$ is that coupling 
of its non-conserving dynamics with the conserved dynamics of the 
$L$ and $R$ species of particles gives rise to an effective 
pheromone-mediated interaction between the particles of the same 
species. This pheromone-mediated interactions between the $L$ ($R$) 
particles gives to a left-moving (right-moving) cluster. This 
tendency for ``coarsening'', induced by the pheromones, then competes 
with the ``shredding'' process which occurs during collision of 
oppositely moving clusters in a {\it finite system} with {\it periodic 
boundary conditions}. 

The most surprising finding is a nontrivial and, at first sight, 
counter-intuitive, system size dependence of the spatio-temporal 
organization which, we believe, could be relevant also for other driven 
systems with competing aggregation and disintegration. It would be 
interesting to obtain a more quantitative description of the shredding 
process that allows to quantify this size dependence. Work in this 
direction is currently under progress.

\ack

We thank Martin Burd, Madhav Gadgil, Raghavendra Gadagkar, Alexander John and  
Joachim Krug for enlightening discussions. 

\section*{References}


\begin{thebibliography}{99}
\bibitem{foot} Number of 
species would be four if vacancies are also regarded as a different 
species.
\end{thebibliography}
\end{document}